\documentclass[aps,amsmath,amssymb,pra,twocolumn,a4paper,reprint,eprint,longbibliography,footinbib]{revtex4-2}
\usepackage{eurosym}
\usepackage[utf8]{inputenc}
\usepackage{graphicx}
\usepackage{amsmath}
\usepackage{amsfonts}
\usepackage{amssymb}
\usepackage{bbm}
\usepackage{bm}
\usepackage{bbold}
\usepackage[usenames,dvipsnames,svgnames,table]{xcolor}
\usepackage[USenglish]{babel}
\usepackage{multirow}
\usepackage{makecell}
\usepackage[colorlinks,bookmarks=false,
            linkcolor=blue,
            anchorcolor=blue,
            citecolor=blue,
            ]{hyperref}
\newcommand{\clr}{\color{red}}

\newcommand{\ignore}[1]{}
\newcommand{\Tr}{\ensuremath{\operatorname{Tr}}}
\usepackage{subfigure}

\begin{document}

\title{The role of superconducting fitness in pairing from fluctuating order}

\author{Yufei Zhu}
\affiliation{Department of Physics and MacDiarmid Institute for Advanced Materials and Nanotechnology, University of Otago, P.O. Box 56, Dunedin 9054, New Zealand}
\author{P. M. R. Brydon}
 \affiliation{Department of Physics and MacDiarmid Institute for Advanced Materials and Nanotechnology,
University of Otago, P.O. Box 56, Dunedin 9054, New Zealand}
\date{\today}

\begin{abstract}
In many unconventional superconductors the pairing interaction is believed to be mediated by a fluctuating order. Although this is typically taken to be magnetic in origin, the role of other fluctuating orders has recently been considered. In this work we examine the weak-coupling pairing interaction produced by a general fluctuating order, and seek to identify the leading pairing instability. For a given pairing channel, we show that the superconducting fitness with the associated static order appears prominently in the expression for the coupling constant. We consequently argue that fit gaps (for which the static order is not pair-breaking) should have an attractive interaction, whereas unfit gaps (for which the static order is pair-breaking) have a repulsive interaction. We propose a simple heuristic test for the tendency of a given pairing state to have an attractive interaction. We show the validity of this test in the case of pairing caused by fluctuating density-wave order, and use it to probe the superconducting state generated by a fluctuating noncolinear magnetic order on the pyrochlore lattice.
\end{abstract}

\maketitle

\section{Introduction}
The origin of the pairing interaction in unconventional superconductors is an important open problem in condensed matter physics. Many unconventional superconducting phases are found upon suppressing a magnetically-ordered state, e.g. by doping or pressure, which is interpreted as evidence in favour of a pairing mechanism based on the exchange of spin-fluctuations~\cite{Berk1966,Rietschel1979,Scalapino1986,Miyake1986,Monthoux1991,Monthoux1992,Scalapino1999,Yanase2003}.
It is both observed in experiment and predicted by theory that certain types of spin fluctuations are associated with particular pairing states, e.g. ferromagnetic spin fluctuations stabilize spin-triplet states, which may be relevant to Uranium-based superconductors~\cite{Taisuke2014,Dai2019}, whereas antiferromagnetic spin fluctuations typically drive sign-changing spin-singlet pairing states, such as the $d_{x^2-y^2}$-wave state in high-$T_{c}$ cuprates and $\mathrm{CeCoIn_5}$~\cite{Miyake1986,Moriya2000,Yanase2003}, or $s_{+-}$-wave pairing in the iron pnictides~\cite{Scalapino2012,Stewart2011,Kuroki2008,Mazin2008}.

Recently there has been interest in the possibility that other types of fluctuating order can mediate unconventional pairing~\cite{Brydon2014,Kozii2015,NomotoPRB2016,Sumita2020,Wang2021,Palle2024}.
For example, parity fluctuations close to a phase that breaks inversion symmetry have been shown to stabilize odd-parity pairing~\cite{Kozii2015,Wang2016}, which may be relevant to Cu$_x$Bi$_2$Se$_3$~\cite{Brydon2014,Wu2017}, Cd$_{2}$Re$_{2}$O$_{7}$~\cite{Harter2017}, and SrTiO$_{3}$~\cite{Schumann2020}. In particular, while odd-parity fluctuations always favour spin-singlet pairing, they also favour certain types of odd-parity pairing, specifically an odd-parity pairing state with pairing potential which is identical to the (static) odd-parity order~\cite{Kozii2015}. Intriguingly, this is exactly the odd-parity pairing state for which the static odd-parity order is not pair-breaking~\cite{Frigeri2004}.

The pair-breaking effect of disorder, inhomogeneity, or external fields is a defining feature of superconductivity and distinguishes different classes of superconductors~\cite{Anderson1961}. This has recently been generalized by the concept of the superconducting fitness~\cite{Ramires2016,Ramires2019}, which quantifies pair-breaking in terms of the degree of interband pairing that a perturbation to the Hamiltonian generates. A remarkable feature of the superconducting fitness is its simplicity, involving only a commutator-like quantity between the pairing potential and the Hamiltonian of the perturbation. This has recently been generalized to encompass the effect of disorder~\cite{Cavanagh2021,Zhu2023} and clean systems where the perturbation is modulated by the wavefunctions of the unperturbed Hamiltonian~\cite{Cavanagh2023}. 

In this work we extend the concept of superconducting fitness to determine the leading pairing instabilities produced by a fluctuation-mediated interaction. Our analysis reveals the intuitively-appealing result that the pairing interaction is attractive (repulsive) for superconducting states which are fit (unfit) with respect to the static order. This result is rigorously established in several cases; More generally, our fitness argument acts as a convenient heuristic test to identify the pairing states favoured by a given fluctuating order. We present two tests of our approach. We first demonstrate the convenience of our approach by considering the well-known case of pairing induced by a fluctuating density-wave order. We further examine the pairing instability in a Luttinger semimetal induced by a fluctuating magnetic order on the pyrochlore lattice. Here we find that our theory successfully identifies the symmetry of the pairing state even when confronted with the complications induced by multiple distinct pairing states in each irrep. 

Our paper is organized as follows. In Sec.~\ref{sec: pairing} we study the pairing interaction within a weak-coupling theory, and show that the superconducting fitness naturally appears in the coupling constant for a given channel. The influence of the form of the interaction potential and basis states for the gap function are discussed in detail in Sec.~\ref{sec:Lorentzian} and~\ref{sec: general basis}, respectively. Further, regarding a phenomenological Lorentzian form of the interaction potential, we argue that this should hold for a general range and thus we introduce a test of the pairing instability in Sec.~\ref{sec:test}. We apply our theory to pairing from fluctuating density waves in Sec.~\ref{sec: density-wave}, and fluctuating magnetic order on the pyrochlore lattice in Sec.~\ref{sec: j=3/2}.

\section{Pairing from fluctuating order}\label{sec: pairing}

Following Refs.~\cite{Kozii2015,Sumita2020}, our starting point is a Hermitian operator describing the coupling of the static order to electrons 
\begin{equation}
  \hat{Q} =  \sum_{\boldsymbol{k}}\sum_{\alpha, \beta}\Gamma_{\boldsymbol{k},\alpha\beta} c_{\boldsymbol{k}, \alpha}^{\dagger} c_{\boldsymbol{k}, \beta},
\end{equation}
where $c_{\boldsymbol{k},\alpha}$ ($c^\dagger_{\boldsymbol{k},\alpha}$) is the electronic annihilation (creation) operator with momentum ${\boldsymbol{k}}$, $\alpha$ and $\beta$ represents a pseudospin degree of freedom labeling doubly-degenerate states of the normal state Hamiltonian. Hermiticity enforces the condition on the vertex function $\Gamma_{\boldsymbol{k},\alpha\beta} = \Gamma^\ast_{\boldsymbol{k},\beta\alpha}$. This can be generalized to a general spatial fluctuation of the static order with wavevector ${\bf q}$ as 
\begin{equation}\label{eq: order}
\hat{\mathcal{Q}}(\boldsymbol{q})=\frac{1}{2} \sum_{\boldsymbol{k}} \sum_{\alpha, \beta}\left\{\Gamma_{\boldsymbol{k}+\boldsymbol{q},\alpha\beta}+\Gamma_{\boldsymbol{k},\alpha\beta}\right\} c_{\boldsymbol{k}+\boldsymbol{q}, \alpha}^{\dagger} c_{\boldsymbol{k}, \beta},
\end{equation}
Integrating out the fluctuating order we obtain the electron-electron interaction
\begin{equation}\label{eq: interaction}
\hat{\mathcal{H}}_{\mathrm{eff}}=\sum_{\boldsymbol{q}} V_{\boldsymbol{q}} \hat{\mathcal{Q}}(\boldsymbol{q}) \hat{\mathcal{Q}}(-\boldsymbol{q}),
\end{equation}
where $V_{\boldsymbol{q}}$ is proportional to the static susceptibility, 
which becomes sharply peaked near the ordering wavevector ${\bf Q}$ as one approaches the instability. 
Restricting the effective interaction Eq.~(\ref{eq: interaction}) to the Cooper channel with zero center-of-mass momentum, we obtain the pairing interaction
\begin{equation}
{\mathcal{H}}_{\mathrm{p}}=
\sum_{\boldsymbol{k}, \boldsymbol{k}^{\prime}} V_{\alpha \beta \gamma \delta}\left(\boldsymbol{k}, \boldsymbol{k}^{\prime}\right) c_{\boldsymbol{k} \alpha}^{\dagger} c_{-\boldsymbol{k} \beta}^{\dagger} c_{-\boldsymbol{k}^{\prime} \gamma} c_{\boldsymbol{k}^{\prime} \delta}, \label{eq:pairingint}
\end{equation}
where the interaction vertex $V_{\alpha \beta \gamma \delta}\left(\boldsymbol{k}, \boldsymbol{k}^{\prime}\right)$ is expressed in terms of the fluctuating order as
\begin{align}
V_{\alpha \beta \gamma \delta}\left(\boldsymbol{k}, \boldsymbol{k}^{\prime}\right)=&\frac{1}{4} V_{\boldsymbol{k}-\boldsymbol{k}^{\prime}}\left(\Gamma_{\boldsymbol{k}}+\Gamma_{\boldsymbol{k}^{\prime}}\right)_{\alpha \delta}\left(\Gamma_{-\boldsymbol{k}}+\Gamma_{-\boldsymbol{k}^{\prime}}\right)_{\beta \gamma}\notag \\
& \times\Theta(\omega_C-|\xi_{\boldsymbol{k}}|)\Theta(\omega_C-|\xi_{\boldsymbol{k}^\prime}|)\label{vertex}
\end{align}
where we have introduced a cutoff $\omega_C\ll \mu$ for the interaction, restricting it to the vicinity of the Fermi surface $\xi_{\boldsymbol{k}}=\epsilon_{\boldsymbol{k}}-\mu = 0$, with $\epsilon_{\boldsymbol{k}}$ the dispersion relation of the electrons and $\mu$ is the chemical potential. Because of the cutoff, it is convenient to restrict the momentum sum in Eq.~(\ref{eq:pairingint}) to states at the Fermi surface, which we emphasize in the following by replacing $\sum_{\boldsymbol{k}}\ldots$ by $\langle \ldots \rangle_{\boldsymbol{k},\text{FS}}$. 

The pairing interaction is decomposed into different Cooper channels as
\begin{equation}
    \hat{\mathcal{H}}_{\mathrm{p}} = \sum_{n} a_{n} \hat{\Delta}_n \hat{\Delta}^\dagger_n \label{diag_pair}
\end{equation}
where the coefficients $a_n$ are the interaction strengths in the various channels, with $a_n<0$ and $a_n>0$ corresponding to an attractive and repulsive channel, respectively, and we define the Cooper pair operator
\begin{equation}
    {\Delta}_{n} = \Bigg\langle\sum_{\alpha\beta}c^\dagger_{{\boldsymbol{k}},\alpha}\Delta_{n,{\boldsymbol{k}},\alpha\beta} c^\dagger_{-\boldsymbol{k},\beta}\Bigg\rangle_{\boldsymbol{k},\text{FS}}
\end{equation}
for the channel $n$ with pairing potential $\hat{\Delta}_{n,\boldsymbol{k}}$. The pairing potentials obey the orthonormal condition
\begin{equation}
    \langle 
\Tr\{\Delta_{n,{\boldsymbol{k}}}\Delta_{n^{\prime},{\boldsymbol{k}}}^\dagger\}\rangle_{\boldsymbol{k},\text{FS}} = \delta_{n,n^{\prime}}.
\end{equation}
The ``diagonal" representation of the pairing interaction Eq.~(\ref{diag_pair}) is of limited practical relevance, as the specific form of the pairing potentials depends on details of the interaction. Nevertheless, the existence of this formal representation is essential for the following general discussion.

Using the orthonormal condition for the pairing potentials, the interaction strengths in Eq.~(\ref{diag_pair}) are related to the interaction vertex Eq.~(\ref{vertex}) by
\begin{widetext}
\begin{equation}
    a_n =  \frac{1}{4}\Bigg\langle\Bigg\langle\sum_{\alpha,\beta,\gamma,\delta} \Delta_{n, \beta \alpha}^{\dagger}(\boldsymbol{k}) V_{\alpha \beta \gamma \delta}\left(\boldsymbol{k}, \boldsymbol{k}^{\prime}\right) \Delta_{n, \delta\gamma}\left(\boldsymbol{k}^{\prime}\right)\Bigg\rangle_{\boldsymbol{k},\text{FS}}\Bigg\rangle_{\boldsymbol{k}',\text{FS}}.\label{eq: amn}
\end{equation}
Inserting the definition of the interaction vertex and writing $\hat{\Delta}_{n,{\bf k}} = \tilde{\Delta}_{n,{\bf k}}i\hat{\sigma}_y$, where $i\hat{\sigma}_y$ is the unitary part of the time-reversal operator, we find the general formula
\begin{equation}
a_{n} = \frac{\text{sgn}(\mathcal{T}_{\Gamma})}{4}\Bigg\langle\Bigg\langle V_{\boldsymbol{k}-\boldsymbol{k}^\prime}\left(
2\Tr\{  \Gamma_{\boldsymbol{k}} \tilde{\Delta}_{n,\boldsymbol{k}} \Gamma_{\boldsymbol{k}}\tilde{\Delta}_{n,\boldsymbol{k}^\prime} \}+
\frac{1}{2}
\Tr\{ \tilde{F}_{-,n,\boldsymbol{k}} \tilde{F}_{-,n,\boldsymbol{k^{\prime}}}
+\tilde{F}_{+,n,\boldsymbol{k}} \tilde{F}_{+,n,\boldsymbol{k^{\prime}}} \}
\right)\Bigg\rangle_{\boldsymbol{k},\text{FS}}\Bigg\rangle_{\boldsymbol{k}',\text{FS}}.\label{eq: general an}
\end{equation}
\end{widetext}
This is the central result of our manuscript. Here $\text{sgn}(\mathcal{T}_{\Gamma})=+1 \ (-1)$ according as the operator $\hat{Q}$ is even (odd) under time-reversal symmetry $\mathcal{T}$.
An important feature of this formula is the presence of the superconducting fitness functions $\tilde{F}_{\pm,n,\boldsymbol{k}}$~\cite{Ramires2016, Ramires2018} in the second trace. These quantities are defined as the commutator ($-$) or anticommutator ($+$) of $\tilde{\Delta}_{n,\boldsymbol{k}}$ with $\Gamma_{\boldsymbol{k}}$: 
\begin{equation}
\tilde{F}_{\pm,n,\boldsymbol{k}}=\Gamma_{\boldsymbol{k}} \tilde{\Delta}_{n,\boldsymbol{k}} \pm \tilde{\Delta}_{n,\boldsymbol{k}} \Gamma_{\boldsymbol{k}}.\label{eq: fit}
\end{equation}
For a time-reversal symmetric (symmetry-breaking) order $\hat{Q}$, $\tilde{F}_{-,n,\boldsymbol{k}}$ ($\tilde{F}_{+,n,\boldsymbol{k}}$) provides the measure of the degree to which it is pair-breaking for the channel $n$, whereas $\tilde{F}_{+,n,\boldsymbol{k}}$ ($\tilde{F}_{-,n,\boldsymbol{k}}$) provides a measure of the insensitivity of channel $n$ to the order. Thus, the fitness function can be understood as quantifying the ``incompatibility'' or ``compatibility'' between the fluctuating order and the pairing. The appearance of $\tilde{F}_{\pm,n,\boldsymbol{k}}$ in Eq.~(\ref{eq: general an}) suggests a connection between the degree of compatibility and the pairing interaction.

\subsection{Role of the interaction potential}\label{sec:Lorentzian}

This connection is complicated by the first trace inside the brackets, and also the interaction potential $V_{\boldsymbol{k}-\boldsymbol{k}^\prime}$ appearing in Eq.~(\ref{eq: general an}). Focusing on the interaction potential, for illustrative purposes we consider an attractive Lorentzian form 
\begin{equation}
    V_{\boldsymbol{k}-\boldsymbol{k^{\prime}}}=\frac{V_{0}}{1+ {\xi}^{2}(\boldsymbol{k}- \boldsymbol{k^{\prime}})^2}, \label{eq:lorentzian}
\end{equation} 
where $V_0<0$ and $\xi$ is the range of the interaction. Note that we restrict to the case where the ordering wavevector ${\bf Q}=0$; we shall see in the following section that our formalism can be straightforwardly adapted to commensurate ordering vectors. Although the formula Eq.~(\ref{eq: general an}) cannot be further simplified for general $\xi$, it is instructive to consider two extreme limits: an on-site interaction $\xi=0$, which gives a constant interaction in momentum space; and an infinite-range interaction $\xi\rightarrow\infty$, which gives a local interaction in momentum space. The former limit may be physically appropriate for a small Fermi surface with characteristic Fermi length $k_F^{-1}$ much larger than the range of the interaction~\cite{Kozii2015, Sumita2020}; on the other hand, the latter limit can simulate the sharply-peaked interaction close to the instability of the fluctuating order.

\subsubsection{On-site interaction}

In the limit of an on-site interaction $\xi=0$ the expression for the coupling constant can be simplified to
\begin{align}
a_n = &  \frac{V_0\text{sgn}({\cal T}_\Gamma)}{8}\left(\Tr\left\{\langle\tilde{F}_{+,n,\boldsymbol{k}}\rangle^2_{\boldsymbol{k},\text{FS}} +\langle \tilde{F}_{-,n,\boldsymbol{k}}\rangle_{\boldsymbol{k},\text{FS}}^2\right\}\right. \notag \\
&\left.+\Tr\left\{\langle \tilde{F}_{+,n,\boldsymbol{k}}\tilde{F}^{\text{av}}_{+,n,\boldsymbol{k}}\rangle_{\boldsymbol{k},\text{FS}} +\langle \tilde{F}_{-,n,\boldsymbol{k}}\tilde{F}^{\text{av}}_{-,n,\boldsymbol{k}}\rangle_{\boldsymbol{k},\text{FS}}\right\}\right) \label{eq:onsite}
\end{align}
where $\tilde{F}^{\text{av}}_{\pm,n,\boldsymbol{k}}=\Gamma_{\boldsymbol{k}} \langle\tilde{\Delta}_{n,\boldsymbol{k}^\prime}\rangle_{\boldsymbol{k}^\prime,\text{FS}} \pm \langle\tilde{\Delta}_{n,\boldsymbol{k}^\prime} \rangle_{\boldsymbol{k}^\prime,\text{FS}}\Gamma_{\boldsymbol{k}}$ is the fitness function for the average gap, and is therefore only present for $s$-wave pairing states. 

We first consider the result for non-$s$-wave gaps (i.e. $\langle \tilde{\Delta}_{n,\boldsymbol{k}}\rangle_{\boldsymbol{k},\text{FS}}=0$), which are allowed when $\Gamma_{\boldsymbol{k}}$ has nontrivial momentum-dependence. In this case only the first line of Eq.~(\ref{eq:onsite}) is relevant.
Since $\tilde{F}_{+,n,\boldsymbol{k}}$ ($\tilde{F}_{-,n,\boldsymbol{k}}$) is Hermitian (anti-Hermitian), its eigenvalues are purely real (imaginary). It follows that the trace of the squared Fermi-surface-averaged fitness is also positive (negative). Thus, for a time-reversal-symmetric (symmetry-breaking) fluctuating order, the leading pairing instability maximizes $\langle \tilde{F}_{+,n,\boldsymbol{k}}\rangle_{\text{FS}}^2$ ($\langle \tilde{F}_{-,n,\boldsymbol{k}}\rangle_{\text{FS}}^2$). That is, the pairing state is perfectly fit with respect to the fluctuating order. Likewise, the most repulsive state is completely unfit. 
The analysis for the $s$-wave states is less straightforward as the terms on the last line spoil the simplicity of the argument. Nevertheless, for an $s$-wave gap we expect that $\tilde{F}_{\pm,n,\boldsymbol{k}}\approx \tilde{F}^{\text{av}}_{\pm,n,\boldsymbol{k}}$, and so a fit (unfit) $s$-wave gap should be expected to lead to an attractive (repulsive) interaction. 

We thus conclude that in the on-site limit the fluctuating order favours pairing states which are fit with respect to the static order, and vice versa. This agrees with the results of Ref.\cite{Kozii2015}, where the time-reversal-symmetric odd-parity fluctuating order generates attractive interactions for $s$-wave singlet pairing and also triplet pairing with the same symmetry, both of which are completely fit with respect to the fluctuating order.

\subsubsection{Infinite-range interaction}

In the limit of an infinite-range interaction we can approximate the interaction potential as a $\delta$-function, i.e. $V_{\boldsymbol{k}-\boldsymbol{k^{\prime}}}= {V}_0 \delta(\boldsymbol{k}-\boldsymbol{k^{\prime}})$, which dramatically simplifies the expression for the coupling constants
\begin{equation}\label{eq:longrange}
a_n = \frac{{V}_0\text{sgn}({\cal T}_\Gamma)}{4} \langle\Tr\{\tilde{F}_{+,n,\boldsymbol{k}}^2 + \tilde{F}_{-,n,\boldsymbol{k}}^2\}\rangle_{\boldsymbol{k},\text{FS}}.
\end{equation} 
Here the role of the fitness is unambiguous: the coupling constant is directly proportional to the usual measure of the fitness of the pairing state with respect to the static order~\cite{Ramires2018}, and the most fit pairing state is the leading instability.

\subsubsection{General interaction} \label{sec:general}

A general interaction potential can be written as
\begin{equation}
V_{\boldsymbol{k}-\boldsymbol{k'}} = \sum_{m}v_m\phi_m(\boldsymbol{k})\phi_m(\boldsymbol{k'})    
\end{equation}
where the $\{\phi_m(\boldsymbol{k})\}$ are a complete set of orthogonal functions and $v_m$ are real coefficients. Inserting this expansion into Eq.~(\ref{eq: general an}), we find that the coupling constant is given by $a_n = \sum_m a^{(m)}_{n}$ where the $a_{n}^{(m)}$ are identical to Eq.~(\ref{eq:onsite}) with the replacements $V_0 \rightarrow v_m$ and $\tilde{\Delta}_{n,\boldsymbol{k}} \rightarrow \phi_m(\boldsymbol{k})\tilde{\Delta}_{n,\boldsymbol{k}}$. This gives a complicated expression for the overall coupling constant, which depends on the sign of $v_m$ and the form of the basis functions $\phi_m(\boldsymbol{k})$. 

To simplify matters, we consider the case where all the $v_m<0$, as is the case of the Lorentzian. In the equivalent of Eq.~(\ref{eq:onsite}) for $a_{n}^{(m)}$, the first line is positive or negative according as the gap function $\phi_{m}(\boldsymbol{k})\tilde{\Delta}_{n,\boldsymbol{k}}$ is fit or unfit with respect to the fluctuating order.  
In general this depends on the basis function $\phi_{m}(\boldsymbol{k})$, and so leads to an inconclusive result. Moreover, the second line is generally nonzero for any gap function, and the sign of this expression is uncertain.

A firm statement can be made in the important case where the gap function $\tilde{\Delta}_{n,\boldsymbol{k}}$ is perfectly fit or unfit, i.e. one of the fitness functions is vanishing. In these cases the first line is maximally negative or positive, respectively. The second line in Eq.~(\ref{eq:onsite}) again prevents a more general statement, although one of the terms will certainly vanish. Due to the Fermi-surface average of the gap which appears here, the other term will only be nonzero if $\phi_{m}(\boldsymbol{k})\tilde{\Delta}_{n,\boldsymbol{k}}$ has an $s$-wave component. So long as the fitness of the Fermi-surface averaged gap is sufficiently similar to the regular fitness, we expect that a fit (unfit) gap will lead to an attractive (repulsive) contribution from the second line. 

The argument simplifies significantly in the case when $\Gamma_{\boldsymbol{k}}$ is momentum-independent: here the first trace in Eq.~(\ref{eq: general an}) is identical to the second, and so the coupling constant is twice the first row of Eq.~(\ref{eq:onsite}). The coupling constant is then maximally attractive for completely fit gaps, but maximally repulsive for completely unfit gaps.

\subsection{General basis functions} \label{sec: general basis}

In the theory above we have assumed a basis of pairing potentials $\{\Delta_{n,\boldsymbol{k}}\}$ such that the pairing interaction is diagonal. Although appropriate for developing our formalism, this is of limited practical use since it requires that we already know the diagonalizing basis. A more practical approach is to expand the interaction in terms of a set of convenient functions representing basis states of the various irreps, e.g. spherical harmonics or tight-binding functions. Since basis states within the same irrep $\Lambda$ of the point group of the crystal can mix, we must generalize the decomposition of the pairing Hamiltonian as
\begin{equation}
{\cal H}_{p} = \sum_{\Lambda}\sum_{n,m\in \Lambda}a^{(\Lambda)}_{n,m}\hat{\Delta}_{n}\hat{\Delta}_{m}^\dagger
\end{equation}
where the coupling constant $a^{(\Lambda)}_{n,m}$ is defined 
\begin{widetext}
\begin{equation}
a^{(\Lambda)}_{n,m} = \frac{\text{sgn}(\mathcal{T}_{\Gamma})}{4}\Bigg\langle\Bigg\langle V_{\boldsymbol{k}-\boldsymbol{k}^\prime}\left(
2\Tr\{  \Gamma_{\boldsymbol{k}} \tilde{\Delta}_{n,\boldsymbol{k}} \Gamma_{\boldsymbol{k}}\tilde{\Delta}_{m,\boldsymbol{k}^\prime} \}+
\frac{1}{2}
\Tr\{ \tilde{F}_{-,n,\boldsymbol{k}} \tilde{F}_{-,m,\boldsymbol{k^{\prime}}}
+\tilde{F}_{+,n,\boldsymbol{k}} \tilde{F}_{+,m,\boldsymbol{k^{\prime}}} \}
\right)\Bigg\rangle_{\boldsymbol{k},\text{FS}}\Bigg\rangle_{\boldsymbol{k}',\text{FS}}.\label{eq: general amn}
\end{equation}
\end{widetext}
To find the leading instability within each irrep, we expand the gap as $\Delta_{\boldsymbol{k}} = \sum_{m\in\Lambda}\Delta^{(m)}_{0}\tilde{\Delta}_{m,\boldsymbol{k}}$. The coefficients $\Delta^{(m)}_0$ obey the linearized gap equation
\begin{equation}\label{eq: linearized GE2}
\Delta_{0}^{(m)}=-\chi(T_{c}) \sum_{l,n\in\Lambda} a^{(\Lambda)}_{m n}  \mathcal{F}^{nl}\Delta_{0}^{(l)},
\end{equation}
where 
$\mathcal{F}^{n l}=\langle \text{Tr}\{\tilde{\Delta}^{\dagger}_{n, \boldsymbol{k}} \tilde{\Delta}_{l, \boldsymbol{k}} \}\rangle_{\boldsymbol{k},\text{FS}}$
denotes the overlap between different components of the pairing state and $\chi(T) = N_0\log(1.14\omega_C/k_BT)$ is the pairing susceptibility. The critical temperature is the highest temperature at which at which \begin{equation}
\det(\mathbb{1}+\chi(T_{c}) \hat{a}^{\Lambda}\hat{\mathcal{F}}) = 0 \label{eq:linearized}
\end{equation}
where $\mathbb{1}$ is the unit matrix, and the elements of the matrices $\hat{a}^\Lambda$ and $\hat{\mathcal{F}}$ are given by $a^{\Lambda}_{m n}$ and $\mathcal{F}^{nl}$, respectively.

The expressions Eq.s~(\ref{eq: general an}) and~(\ref{eq:linearized}) show the fitness concept is of limited utility in this formulation of the problem: the fitness functions of different basis states in the same irrep are generally different, and so it is not possible to use this to make any statement about the off-diagonal coupling constants; moreover, the effect of these off-diagonal coupling constants is modulated by the overlap terms $\mathcal{F}^{nl}$, which are not captured by the fitness argument. The fitness approach nevertheless remains useful if the pairing potential is dominated by a single basis function. This is frequently the case, especially in the case of short-range pairing interactions.

\subsection{Fitness as a heuristic}\label{sec:test}

Although we were only able to make a firm conclusion in limiting cases, the discussion above strongly suggests that the pairing interaction due to a fluctuating order will favor pairing in channels which are fit with respect to the corresponding static order, while being repulsive for the unfit superconducting channels.
To practically apply this insight, we introduce the compatibility measure of the static order with a given pairing state $\Delta_{\boldsymbol{k}}$:
\begin{equation}
 \mathbb{F} = \text{sgn}({\cal T}_\Gamma) \langle\Tr\{\tilde{F}_{+,\boldsymbol{k}}^2 + \tilde{F}_{-,\boldsymbol{k}}^2\}\rangle_{\boldsymbol{k},\text{FS}}\,.\label{eq: fitness argument}
\end{equation}  
The more positive $\mathbb{F}$, the stronger the tendency that the fluctuating order induces an attractive interaction; conversely, the more negative $\mathbb{F}$, the more likely the pairing interaction is repulsive. 
Although of only approximate validity, this quantity is very simple to calculate and can act as a convenient guideline for assessing the pairing near a non-superconducting ordered phase.

\section{Pairing near a density-wave instability \label{sec: density-wave}}

As a first application of our formalism, we consider pairing near a density-wave instability, since unconventional SC phases are often found in proximity to a spin-density wave (SDW) or charge-density wave (CDW) state~\cite{Patton1973,Scalapino1986,Mutka1981,Johannes2008,Kim2016}. Although the theory in Sec.~\ref{sec: pairing} has been developed for fluctuations of a uniform (${\bf Q}=0$) order, it is straightforward to generalize our theory to a fluctuating density-wave order.
We consider a CDW or SDW with wavevector $\boldsymbol{Q}=\frac{1}{2}
\mathbf{G}$, where $\mathbf{G}$ is the reciprocal lattice
vector. The operator for these orders can be expressed as
\begin{align}
\hat{\mathcal{Q}}_{\mathrm{CDW}}=
  &\sideset{}{'}\sum_{\boldsymbol{k}}\Psi_{\boldsymbol{k}}^{\dagger}
    {\Gamma_{\mathrm{CDW}}}\Psi_{\boldsymbol{k}} \\
\hat{\boldsymbol{\mathcal{Q}}}_{\mathrm{SDW}}=& \sideset{}{'}\sum_{\boldsymbol{k}} \Psi_{\boldsymbol{k}}^{\dagger}{\boldsymbol{\Gamma}_{\mathrm{SDW}}} \Psi_{\boldsymbol{k}}
\end{align}
where ${\Gamma_{\mathrm{CDW}}= \tau_x \sigma_{0}}$ and
$\boldsymbol{\Gamma}_{\mathrm{SDW}} =\tau_x\boldsymbol{\sigma}$, the primed sum runs over the reduced Brillouin zone, and
$\Psi_{\boldsymbol{k}}$ is a spinor of annihilation operators defined
\begin{equation}
\Psi_{\boldsymbol{k}}=\left(c_{\boldsymbol{k}, \uparrow}, c_{\boldsymbol{k}, \downarrow}, c_{\boldsymbol{k}+\frac{1}{2}
\boldsymbol{G}, \uparrow}, c_{\boldsymbol{k}+\frac{1}{2}
\boldsymbol{G}, \downarrow}\right)^{T}\,.
\end{equation}
The umklapp terms (defined at $\boldsymbol{k}+\frac{1}{2}
\boldsymbol{G}$) in the definition of $\Psi_{\boldsymbol{k}}$ can be considered as an additional discrete degree of 
freedom, which is encoded by the Pauli matrix
$\tau_{x}$ in the definitions of $\Gamma_{\mathrm{CDW}}$ and $\Gamma_{\mathrm{SDW}}$. Expressed in this basis, the density wave is a ${{\boldsymbol{Q}}}=0$
instability, which immediately allows us to apply the formalism
developed above.

Ignoring the possibility of a pair-density wave, the general expression for a pairing potential in this basis is 
\begin{equation}
\tilde{\Delta} =
\left(\begin{array}{cc}
\tilde{\Delta}_{\boldsymbol{k}} & 0 \\
0 & \tilde{\Delta}_{\boldsymbol{k}+\frac{1}{2}
\boldsymbol{G}}
\end{array}\right).
\end{equation}
We specialize in two limits involving the umklapp degree of freedom:
sign-preserving (``umklapp-trivial") pairing states where
$\tilde{\Delta}_{\boldsymbol{k}}=\tilde{\Delta}_{{\boldsymbol{k}+\frac{1}{2}
\boldsymbol{G}}}$, which implies
trivial $\tau$-dependence, i.e.
$\tilde{\Delta} = \tilde{\Delta}_{\boldsymbol{k}}\tau_0$; and a
$\tilde{\Delta}_{\boldsymbol{k}}=-\tilde{\Delta}_{{\boldsymbol{k}+\frac{1}{2}
\boldsymbol{G}}}$, giving
the umklapp-nontrivial form $\tilde{\Delta} = \tilde{\Delta}_{\boldsymbol{k}}\tau_z$.
We further classify the pairing as spin-singlet
($\tilde{\Delta}_{\boldsymbol{k}}=\Phi_{\boldsymbol{k}}\sigma_{0}$) or
spin-triplet
($\tilde{\Delta}_{\boldsymbol{k}}=\mathbf{d}_{\boldsymbol{k}}\cdot\boldsymbol{\sigma}$). A general pairing state will have both umklapp trivial and nontrivial components.

\begin{table}
\begin{center}
\begin{tabular}{|c|c|c|c|c|}
 \hline \hline  spin & umklapp & $\tilde{\Delta}$  & $\mathbb{F}$ CDW  &  $\mathbb{F}$ $\mu$-SDW \\

\hline \multirow{2}*{singlet} & trivial & $\Phi_{\boldsymbol{k}} \tau_{0}{\sigma}_{0}$ &   $\Phi_{\boldsymbol{k}}^{2}  $ & $-\Phi_{\boldsymbol{k}}^{2}$ \\

~ &nontrivial & $\Phi_{\boldsymbol{k}} \tau_{z} \sigma_{0}$ &   $-\Phi_{\boldsymbol{k}}^{2} $ &  $\Phi_{\boldsymbol{k}}^{2}$  \\
\hline \multirow{2}*{triplet} & trivial & ${\tau}_{0} {{\bf d}_{\boldsymbol{k}}} \cdot\boldsymbol{\sigma}$  & $|\mathbf{d}_{\boldsymbol{k}}|^{2} $ &  $| \mathbf{d}_{\boldsymbol{k}}|^{2}-2(d^\mu_{\boldsymbol{k}})^{2}$ \\
~& nontrivial &${\tau}_{z} {{\bf d}_{\boldsymbol{k}}} \cdot\boldsymbol{\sigma}$ &  $-|\mathbf{d}_{\boldsymbol{k}}|^{2} $ &  $2(d^\mu_{\boldsymbol{k}})^{2}-|\mathbf{d}_{\boldsymbol{k}}|^{2}$ \\
\hline \hline
\end{tabular}
\end{center}
\caption{The compatibility measure $\mathbb{F}$ in Eq.~(\ref{eq: fitness argument}) for possible pairing states due to fluctuating CDW or a $\mu$-component SDW order.}\label{table1}
\end{table}

In Table \ref{table1} we present the results of evaluating
Eq.~(\ref{eq: fitness argument}) for the different gap structures and fluctuating
order. Examining first the spin-singlet states, we observe that
fluctuating CDW order favours attractive interactions for umklapp-trivial pairing.
In contrast, the compatibility measure for a fluctuating SDW order favours attractive interactions for umklapp-nontrivial pairing
states. This is consistent with the results of more
sophisticated treatments: Taking the most famous example, for a two-dimensional square
lattice with $(\pi,\pi)$-fluctuating magnetic order, as is relevant to the cuprates \cite{Miyake1986,Yanase2003,Scalapino2012},
our theory implies attractive interactions for nearest-neighbour extended $s$-wave and $d_{x^2-y^2}$-wave
pairing, whereas next-nearest-neighbour $d_{xy}$-wave states are
repulsive.

Turning now to the triplet channel, we observe that the fluctuating CDW order again only gives attractive interactions for an umklapp-trivial pairing state. The situation for the fluctuating SDW order depends on the spin polarization of the SDW~\cite{Ozaki1989}: a fluctuating $\mu$-component will favor a umklapp-trivial pairing state with ${\bf d}$-vector perpendicular to the $\mu$-axis, and a umklapp-nontrivial pairing state with ${\bf d}$-vector parallel to the $\mu$-axis. For spin-isotropic SDW fluctuations, we expect only to find an attractive pairing for umklapp-trivial states. In the context of the cuprates, where the spin-orbit coupling is weak and the SDW fluctuations are isotropic, this excludes nearest-neighbour triplet pairing states ($\propto \sin(k_xa)$, etc) which change sign under $\boldsymbol{k} \to {\boldsymbol{k}+\frac{1}{2} \boldsymbol{G}}$. Although umklapp-trivial next-nearest-neighbour triplet pairing is possible, the pairing strength is likely to be much weaker than for the nearest-neighbour singlet pairing since we expect the interaction potential to drop off quickly with distance.

\begin{table*}
\def\arraystretch{1.6}
\scalebox{1.0}{
\begin{tabular}{|c|c|c|c|c|c|c|c|c|}
\hline  \multirow{2}*{\text {irrep of  $T_d$}} & \multirow{2}*{ $\tilde{\Delta}\left(\boldsymbol{k}\right)$} & \multicolumn{3}{c|}{$m_j=1/2$} & \multicolumn{3}{c|}{$m_j=3/2$} \\
\cline{3-8} 
& &   $\mathbb{F}$ & $\mathcal{F}$ 
&  $\tilde{\Delta}_{\mathbb{F}}$ &   $\mathbb{F}$ 
& $\mathcal{F}$ &  $\tilde{\Delta}_{\mathbb{F}}$ \\
\hline 
$A_{1}$ & $\frac{1}{\sqrt{3}} \mathbf{T} \cdot \boldsymbol{k} $&2.120 & 0.171 & 0.363 & 2.360 & 2.571 & {\clr 6.068 }\\
\hline
\multirow{2}*{$A_{2}$} & $\frac{2}{\sqrt{5}}\mathbf{J}\cdot \boldsymbol{k} $ &0.288 & 0.400 & 0.115 & 1.808 & 3.600 & {\clr 6.509 }\\
~ & $\boldsymbol{\mathcal{J}}\cdot \boldsymbol{k} $ & -0.140 & 1.543 & -0.216 & -1.216 & 0.743 & -0.903 \\
\hline 
\multirow{3}*{$E$}  & $\frac{\sqrt{2}}{\sqrt{5}}\left[-J_{x} k_{x}-J_{y} k_{y}+2 J_{z} k_{z}, \sqrt{3}\left(J_{x} k_{x}-J_{y} k_{y}\right)\right]$ &2.136 & 1.120 & {\clr 2.392} & 1.344 & 1.440 & 1.935\\
~ & $\frac{1}{\sqrt{2}}\left[-\mathcal{J}_{x} k_{x}-\mathcal{J}_{y} k_{y}+2 \mathcal{J}_{z} k_{z}, \sqrt{3}\left(\mathcal{J}_{x} k_{x}-\mathcal{J}_{y} k_{y}\right)\right]$ &1.600 & 1.851 & {\clr 2.962} & 1.696 & 0.331 & 0.561 \\
~& $\frac{1}{\sqrt{6}}\left[T_{x} k_{x}+T_{y} k_{y}-2 T_{z} k_{z}, \sqrt{3}\left(T_{x} k_{x}-T_{y} k_{y}\right)\right]$  &2.216 & 1.714 & {\clr 3.798} & 1.816 & 0.514 & 0.933 \\
\hline 
\multirow{4}*{$T_{1}$} & $\frac{\sqrt{6}}{\sqrt{5}}\left(J_{y} k_{z}+J_{z} k_{y}, J_{x} k_{z}+J_{z} k_{x}, J_{x} k_{y}+J_{y} k_{x}\right)$ &-0.128 & 1.120 & -0.143 & 2.120 & 1.440 & 3.053\\
~ & $\frac{4}{\sqrt{3}}(J_xJ_yJ_z + J_zJ_yJ_x)\boldsymbol{k}$ & -2.380 & 1.600 & -3.808 & -2.380 & 1.600 & -3.808 \\
~ & $\frac{\sqrt{3}}{\sqrt{2}}\left(\mathcal{J}_{y} k_{z}+\mathcal{J}_{z} k_{y}, \mathcal{J}_{x} k_{z}+\mathcal{J}_{z} k_{x}, \mathcal{J}_{x} k_{y}+\mathcal{J}_{y} k_{x}\right)$&-1.400 & 0.823 & -1.152 & -0.880 & 1.703 & -1.499\\
~ & $\frac{1}{\sqrt{2}} \mathbf{T} \times \boldsymbol{k}$ & 1.728 & 1.714 & 2.962 &0.200 & 0.514 & 0.103\\
\hline 
\multirow{3}*{$T_{2}$} & $ \frac{\sqrt{6}}{\sqrt{5}}\mathbf{J} \times \boldsymbol{k}$ & 0.856 & 1.600 & 1.370 & $\emptyset$ & 0 & $\emptyset$\\
~ & $ \frac{\sqrt{3}}{\sqrt{2}}\boldsymbol{\mathcal{J}} \times \boldsymbol{k}$ & 0.936 & 1.029 & 0.963 & -0.360 & 1.429 & -0.514\\
~ & $\frac{1}{\sqrt{2}}\left(T_{y} k_{z}+T_{z} k_{y}, T_{x} k_{z}+T_{z} k_{x}, T_{x} k_{y}+T_{y} k_{x}\right)$ &0.984 & 0.686 & 0.675 & 1.872 & 1.886 & 3.531 \\
\hline
\end{tabular}}
\caption{Evaluations of Eq.~(\ref{eq: fitness argument}), the gap magnitude, and the product of them of odd-parity unconventional superconducting states in the point group $T_{d}$ when the Fermi energy sets at the $j=1/2$ band and at the $j=3/2$ band.
The first column: all irreducible odd parity representations in the $T_{d}$ group.
The second column: pairing matrices $\tilde{\Delta}_{\boldsymbol{k}}$ up to $p$-wave under the $j=3/2$ representation. 
Red numbers: the most likely favored pairing states according to the most positive values of $\tilde{\Delta}_{\mathbb{F}} = \mathbb{F}\mathcal{F}$.
Notations: We define $\boldsymbol{\mathcal{J}}=\frac{-41}{6 \sqrt{5}} \mathbf{J}+\frac{2 \sqrt{5}}{3}\left(J_{x}^{3}, J_{y}^{3}, J_{z}^{3}\right)$, $T_{x}=\left\{J_{x}, J_{y}^{2}-J_{z}^{2}\right\}$ and $T_{y,z}$ are given by cyclic permutations.
}\label{tab: AIAO}
\end{table*}

\section{AIAO order in Luttinger semimetals \label{sec: j=3/2}}

As a second example we consider pairing from fluctuating magnetic order on a pyrochlore lattice. 
The pyrochlore lattice consists of the corner-sharing tetrahedron, and is realized in a number of strongly-correlated materials, e.g. the location of the iridium ions in the rare-earth iridates $R_2$Ir$_2$O$_7$ ($R$ is a rare-earth element). Due to the magnetic moments of the iridium ions and the inherent frustration of the pyrochlore lattice, these systems have been predicted to display so-called all-in-all-out (AIAO) order where the magnetic moments on each tetrahedron either point towards or away from the tetrahedron's centre~\cite{Wan2011,WitczakKrempa2012, WitczakKrempa2013, Yang2014, Goswami2017, Berke2018}; this state has been confirmed in several family members~\cite{Ueda2015,Sushkov2015,Donnerer2016,Tomiyasu2012}. 
An intriguing aspect of these systems is that the low-energy fermiology is described by a Luttinger semimetal~\cite{Kondo2015,Nakayama2016}, where strong spin-orbit coupling leads to an effective $j=3/2$ angular momentum of the electrons. Doping the pyrochlore lattice suppresses the AIAO order~\cite{Porter2019,Shinaoka2015}, while also causing the appearance of a Fermi surface. The superconducting states of $j=3/2$ electrons have attracted much attention recently~\cite{Brydon2016,Savary2017,Boettcher2018,Venderbos2018,Link2022}, and so in this section we explore the interplay of the fluctuating order and the spin-orbit texture of the band states control the leading pairing instability.

\subsection{Effective $j=3/2$ electrons}

Up to the quadratic order in $\boldsymbol{k}$, the $j = 3/2$ electron states at the centre of the Brillouin zone are described by the Luttinger-Kohn model~\cite{LuttingerKohn1955} with Hamiltonian matrix
\begin{equation}
\begin{aligned}
\mathcal{H}_{0}=& \left(\beta_{0} \boldsymbol{|k|}^{2}-\mu\right) +\beta_{1}\sum_{i}k_{i}^{2}J_{i}
+\beta_{2}\sum_{i^{\prime}\ne i}k_{i}k_{i^{\prime}}J_{i}J_{i^{\prime}}, \label{eq:LKmodel}
\end{aligned}
\end{equation}
where $J_{i=x,y,z}$ are $4 \times 4$ angular momentum matrices of $j=3/2$ electrons and $\mu$ is the chemical potential. 
The $\beta_{i=0,1,2}$ are material-dependent parameters determined by the electronic band structure; notably, $\beta_{i=1,2}$ are the symmetric spin-orbit coupling terms. For simplicity, we consider the spherical limit with $\beta_{1}=\beta_{2}$; we do not expect our results to significantly change for small cubic anisotropy. The energy eigenvalues of Eq.~(\ref{eq:LKmodel}) are then given by
\begin{equation}
    E_{\pm,\boldsymbol{k}} = \left(\beta_0 + \frac{5}{4}\beta_1\pm \beta_1\right)\boldsymbol{|k|}^{2} - \mu
\end{equation}
which describes two parabolic bands which touch at the $\Gamma$ point. A Luttinger semimetal is realized when the two bands have opposite curvature, which for positive $\beta_0$ is realized when $-9\beta_1>4\beta_0>-\beta_1$. The band-touching point is located at chemical potential $\mu=0$; doping gives a nonzero $\mu$, such that a Fermi surface is realized in either the upper or lower bands. In the spherical limit $(\boldsymbol{k}\cdot\mathbf{J})^2$ is a good quantum number, and so we can label the bands as either $m_j=\frac{3}{2}$ ($E_{+,\boldsymbol{k}}$) or $m_j=\frac{1}{2}$ ($E_{-,\boldsymbol{k}}$), corresponding to $(\boldsymbol{k}\cdot\mathbf{J})^2 = \frac{9}{4}|\boldsymbol{k}|^2$) or $ \frac{1}{4}|\boldsymbol{k}|^2$, respectively.

\subsection{Pairing channels}

In the $j=3/2$ representation, the AIAO order couples to the low-energy electron states with the momentum-independent form
\begin{equation}
\Gamma=J_{x} J_{y} J_{z}+J_{z} J_{y} J_{x}.
\end{equation}
However, we cannot directly evaluate the compatibility measure using this form, since we must first restrict to states at the Fermi energy, corresponding either to the $m_j=\frac{3}{2}$ or $m_j=\frac{1}{2}$ bands. To do this, we must project $\Gamma$ and the gap function into the low-energy subspace, i.e. $\Gamma \rightarrow \Gamma_{m_j,\boldsymbol{k}}= {\cal P}_{m_j,\boldsymbol{k}}\Gamma{\cal P}_{m_j,\boldsymbol{k}}$, where ${\cal P}_{m_j,\boldsymbol{k}} = [(\hat{\boldsymbol{k}}\cdot\mathbf{J})^2-\overline{m}_j^2]/(m_j^2-\overline{m}_j^2)$ is the projection operator onto the $m_j$ band, and $\overline{m}_j=\frac{1}{2}$ for $m_j=\frac{3}{2}$ and \emph{vice versa}. The projection operation is nontrivial,  and typically introduces a momentum-dependence to the vertex for the fluctuating order; it is well known that the projection can produce nodal structures into the superconducting gap.

\begin{figure}[t!]
\centering
\begin{subfigure}
  \centering
  \includegraphics[width=0.80\linewidth]{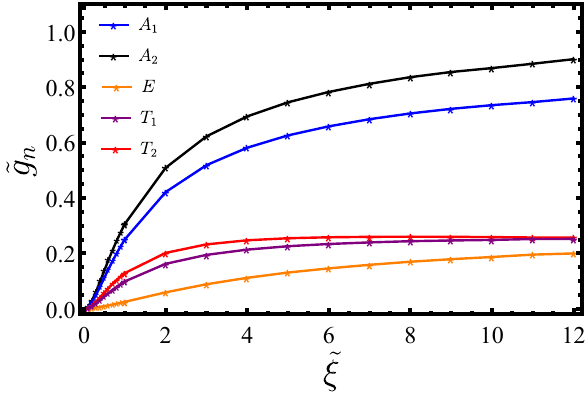}
  \put(-200, 10){{(a)}}
\end{subfigure}%
\\
\begin{subfigure}
  \centering
  \includegraphics[width=0.80\linewidth]{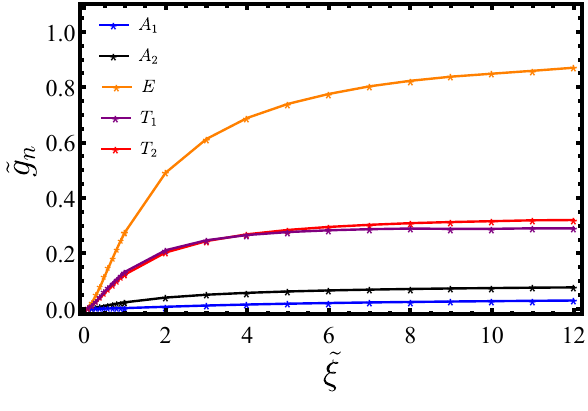}
  \put(-200, 10){{(b)}}
\end{subfigure}
\caption{The plot of maximum eigenvalue $\tilde{g}_{n}$ of each $p$-wave state in the $T_{d}$ group as a function of $\tilde{\xi}=\xi k_F$ at the (a) $m_j=3/2$ Fermi surface (b) $m_j=1/2$ Fermi surface. }
\label{fig:lorentzian}
\end{figure}

The fluctuating AIAO order introduces repulsive interactions in all even-parity channels. This follows from the projection into the low-energy subspace, which due to inversion and time-reversal symmetry can be formulated in terms of a pseudospin basis. Within this basis the AIAO order has the form $\Gamma_{m_j,\boldsymbol{k}} = \sum_{\nu=x,y,z}\epsilon_{\nu,m_j,\boldsymbol{k}}\hat{\varsigma}^\nu$ where the $\hat{\varsigma}^\nu$ are pseudospin Pauli matrices and the $\epsilon_{\nu,m_j,\boldsymbol{k}}$ are momentum-even coefficients. Since the even-parity states are pseudospin-singlet, we must have $\tilde{\Delta}_{\boldsymbol{k}}\propto \hat{\varsigma}^0$. As such, the even-parity states are completely unfit with respect to the AIAO order, and so from the discussion in Sec.~\ref{sec:general} it follows that the pairing attraction is repulsive.

Turning to the odd-parity pairing channels, we restrict our attention to $p$-wave pairing states due to the small size of the Fermi surface in a doped semimetal. In Table~\ref{tab: AIAO} we list all the odd-parity $p$-wave pairing states for $j=3/2$ fermions, classified in terms of the irreps of the $T_d$ point group~\cite{Savary2017}. With the exception of the $A_1$ irrep, each irrep contains multiple distinct states, and so as discussed in Sec.~\ref{sec: general basis} we are unable to rigorously relate the fitness to attractive interactions in the various channels. An added complication here is that the projection of the odd-parity states onto the Fermi surface is nontrivial, and can introduce additional gap nodes not present in the bare potential~\cite{Savary2017}. As a measure of the projected gap we introduce the quantity $\mathcal{F}=\langle \text{Tr}\{\tilde{\Delta}^{\dagger}_{\boldsymbol{k}} \mathcal{P}_{m_j\boldsymbol{k}} \tilde{\Delta}_{\boldsymbol{k}} \mathcal{P}_{m_j\boldsymbol{k}} \}\rangle_{\boldsymbol{k},\text{FS}}$. 
Based on the reasoning in Sec.~\ref{sec: general basis}, our hypothesis is that when ignoring the off-diagonal coupling, the product of this quantity and the compatibility measure should be maximized by the leading pairing instability; as such, it is convenient to introduce 
\begin{equation}\label{eq: fit gap}
    \tilde{\Delta}_{\mathbb{F}}= \mathbb{F}\times \mathcal{F},
\end{equation}
which should be maximal for the leading pairing state.

In Table~\ref{tab: AIAO}, we present the three quantities $\mathbb{F}$, $\mathcal{F}$, and $\tilde{\Delta}_{\mathbb{F}}$ for each of the $p$-wave pairing states. Note that the values on the $m_j=1/2$ and $m_j=3/2$ Fermi surfaces are quite different, showing the nontrivial effect of the Fermi-surface projection. This is also apparent in the maximal values of $\tilde{\Delta}_{\mathbb{F}}$: on the $m_j=1/2$ Fermi surface the three states in the $E$ irrep all have relatively large positive values, while on the $m_j=3/2$ Fermi surface the largest values are found for states are in the $A_2$ and $A_1$ irreps. It is interesting to note the interplay of the quantities $\mathbb{F}$ and $\mathcal{F}$: at the $m_j=3/2$ Fermi surface the $A_1$ state has the highest compatibility measure, but the larger value of $\mathcal{F}$ for the first $A_2$ state gives the larger value of $\tilde{\Delta}_{\mathbb{F}}$. Note that $\mathbb{F}$ of the first $T_2$ channel at the $m_j=3/2$ Fermi surface is ill-defined because the gap magnitude is vanishing, i.e. this state does not open a gap on the $m_j=3/2$ Fermi surface. 
On the basis of the ``rule of thumb" analysis of $\tilde{\Delta}_{\mathbb{F}}$, we expect that $E$ states will be favoured by the AIAO order on the $m_j=1/2$ Fermi surface, while there will be a close competition between $A_1$ and $A_2$ states on the $m_j=3/2$ Fermi surface.

In order to check our results we solve the linearized gap equation including the coupling between the different basis states. Assuming a Lorentzian interaction potential Eq.~(\ref{eq:lorentzian}), we evaluate the $a_{mn}$ according to Eq.~(\ref{eq: general an}). To solve the linearized gap equation~(\ref{eq: linearized GE2}), we find the eigenvalues $\{g_n\}$ of the matrix $\hat{a}^\Lambda\hat{\mathcal{F}}$; the largest eigenvalue $\tilde{g}_n=\text{Max}\{g_{n}/V_{0}\}$ gives the leading instability, with $T_c$ determined by $1 + \tilde{g}_nV_{0}\chi(T_c)=0$. We plot $\tilde{g}_n$ for each irrep as a function of the dimensionless Lorentzian range $\tilde{\xi}=\xi k_F$ in Fig.~\ref{fig:lorentzian}; the irrep with the highest value of $\tilde{g}_n$ defines the leading instability.
Our results are in excellent agreement with our interpretation of the $\tilde{\Delta}_{\mathbb{F}}$ in Table~\ref{tab: AIAO}: for all values of $\xi$, the leading instability on the $m_j=1/2$ Fermi surface is in the $E$ irrep, while on the $m_j=3/2$ Fermi surface the leading instability is in the $A_2$ irrep, closely followed by the $A_1$ irrep. 

Our conclusions are in agreement with the analysis of a general phenomenological $p$-wave pairing interaction for the $\beta_0=0$ Luttinger semimetal presented in Ref.~\cite{Link2022}. The authors classified the pairing states in terms of the total $J=L+S$ angular momentum of the Cooper pairs. On the $m_j=3/2$ Fermi surface the authors found competing $A_1$ and triplet $A_2$ (i.e. $\tilde{\Delta}({\bm k}) = \frac{2}{\sqrt{5}}{\bf J}\cdot{\bm k}$) states, which correspond to the states with the largest value of $\mathbb{F}$ in Table~\ref{tab: AIAO}. The situation on the $m_j=1/2$ Fermi surface is more complicated: here the authors found either a triplet $T_2$ state (i.e. $\tilde{\Delta}({\bm k}) = \frac{\sqrt{6}}{\sqrt{5}}{\bf J}\times{\bm k}$), or $J=2$ states which fall into either the $E$ or $T_1$ irreps. The $|J,M\rangle = |2,0\rangle$ state, which belongs to the $E$ irrep, dominates the $J=2$ region of the phase diagram. Notably, the $A_2$ irrep is stable on the $m_j=3/2$ Fermi surface for parameters where the $E$ irrep is realized on the $m_j=1/2$ Fermi surface, which is consistent with our findings. We also note the recent microscopic numerical calculation of the RPA pairing interaction for the pyrochlore lattice, where it was claimed that the leading instability belongs to the $T_1$ irrep~\cite{Sumiyoshiya2023}. However, significant cubic anisotropy in their work prevents direct comparison to our results and Ref.~\cite{Link2022}.
The role of the spin-orbit texture in setting the leading instability is a promising direction for future research.

\subsection{Superconductivity in half-Heuslers}

We are not aware of any experimental reports of superconductivity arising from doping a compound displaying AIAO order. However, nodal superconductivity has been realized in doped Luttinger semimetals, specifically the noncentrosymmetric half-Heusler compounds YPtBi~\cite{Kim2018} and LuPdBi~\cite{Ishihara2021}, which realize doped Luttinger semimetals. The nodes have been interpreted as originating from the coexistence of $s$-wave singlet and $p$-wave septet pairing states, with the latter corresponding to the $A_1$ state considered here. The origin of the pairing interaction in these materials remains unclear, but the large $p$-wave component is unlikely to arise from phonons. Although these systems are not known to be close to an ordered state instability, recent $^{195}$Pt-NMR experiments in YPtBi report antiferromagnetic spin fluctuations in the normal state~\cite{Zhou2023}. Although YPtBi does not have the pyrochlore structure, and so the AIAO fluctuations discussed here are not directly applicable, our work nevertheless shows that antiferromagnet-like fluctuations in a Luttinger semimetal can mediate pairing in an odd-parity channel.

\section{Conclusions} \label{sec: conclusions}

In this work we have developed a weak-coupling theory for Cooper pairing mediated by a general fluctuating order. Our chief insight is that whether the pairing interaction is attractive or repulsive in a particular superconducting channel can be related to the fitness of the gap function with respect to the static order; specifically, fit gaps are expected to experience an attractive interaction, whereas the interaction is repulsive for unfit gaps. Although a rigorous connection between fitness and the pairing interaction has proved elusive, the relationship is clear in the limits of an on-site and infinite-range interaction. We consequently propose the comparability measure Eq.~(\ref{eq: fitness argument}) based on the fitness for determining if a given pairing state could be realized by a fluctuating order. 

We have applied our theory to the familiar case of pairing near a density-wave instability, verifying the well-known result that fluctuating charge-density and spin-density waves favour sign-preserving and sign-changing gaps, respectively. Our results are notable in that they were obtained via simple matrix algebra, as opposed to complicated many-body theory calculations. As a test for a case where the pairing state is not known, and may involve a superposition of several possible basis states, we have considered the pairing instability that arises upon doping the AIAO-ordered state on the pyrochlore lattice. We find excellent agreement between the fitness argument and the solution of the linearized gap equation, with states in the $A_1$ and $A_2$ irreps closely competing on the $m_j=3/2$ Fermi surface, while a state in the $E$ irrep is realized on the $m_j=1/2$ Fermi surface.

Our work establishes superconducting fitness as a simple guide to understanding the phase diagram of unconventional superconductors. Although we anticipate that exceptions to the ``rule" may be identified, we believe that it should be generally valid. It is also physically appealing, as the fitness associates repulsive interactions with pair-breaking orders. Our analysis further reveals the deep reach of the fitness concept~\cite{Ramires2016,Ramires2018} in the physics of superconductivity.

\acknowledgments

This work was supported by the Marsden Fund Council from Government funding, managed by Royal Society Te Ap\={a}rangi, Contract No. UOO1836.

\bibliography{AIAO_ref}

\end{document}